\documentclass[hyper]{JHEP} % 10pt is ignored!

\usepackage{epsfig}
\usepackage{graphicx}% Include figure files

\def\be{\begin{equation}}
\def\ee{\end{equation}}
\def\ba{\begin{array}}
\def\ea{\end{array}}
\def\bea{\begin{eqnarray}}
\def\eea{\end{eqnarray}}
\def\nn{\nonumber\\}

\def\ct{\cite}
\def\la{\label}
\def\eq#1{Eq. (\ref{#1})}

\def\a{\alpha}
\def\b{\beta}

\def\G{\Gamma}

\def\D{\Delta}

\def\ph{\phi}

\def\ps{\psi}

\def\k{\kappa}
\def\l{\lambda}

\def\m{\mu}
\def\n{\nu}
\def\th{\theta}

\def\r{\rho}
\def\s{\sigma}

\def\ta{\tau}

\def\pr{\prime}

\def\half{\frac{1}{2}}

\def\pa{\partial}

\def\lb{\left[}
\def\lc{\left\{}
\def\ls{\left(}
\def\lp{\left.}
\def\rp{\right.}
\def\rb{\right]}
\def\rc{\right\}}
\def\rs{\right)}
\def\td{\tilde}

\def\det{{\rm det}}
\def\arccos{{\rm arccos}}

\newcommand\fverb{\setbox\pippobox=\hbox\bgroup\verb}

\newcommand\fverbdo{\egroup\medskip\noindent%

            \fbox{\unhbox\pippobox}\ }

\newcommand\fverbit{\egroup\item[\fbox{\unhbox\pippobox}]}

\newbox\pippobox

\title{Spiky Strings on AdS$_4 \times {\bf CP}^3$}

\author{Bum-Hoon Lee $^a$, Kamal L. Panigrahi $^b$ and Chanyong Park $^{a,c}$\\

\vspace{1cm}

$^a$ Center for Quantum Spacetime (CQUeST), Sogang University, \\
~~Seoul 121-742, Korea \\

$^b$ Department of Physics, Indian Institute of Technology Guwahati,\\
~~Guwahati-781 039, India \\

$^c$  National Institute
for Mathematical Sciences, Daejeon 305-340, Korea  \\

\vspace{1cm}

E-mail: \email{bhl@sogang.ac.kr, panigrahi@iitg.ernet.in, cyong21@sogang.ac.kr}}

\preprint{0807.2559}

 \abstract{We study a giant magnon and a spike solution for the string rotating on
 $AdS_4 \times {\bf CP}^3$ geometry. We consider rigid rotating
 fundamental string in the $SU(2)\times SU(2)$ sector inside the ${\bf CP}^3$ and find out the
 general form of all the conserved charges. We find out the
 dispersion relation corresponding to both the known giant magnon and the new
 spike solutions. We further study the finite size correction in both cases.}

\keywords{AdS-CFT Correspondence}

\begin{document}
%%%%%%%%%%%%%%%%%%%%%%%%%%%%%%%%%%%%%%%%%%
%%%%redukovana verze clanku            %%%%
%%%%o Ispike_a3.tex%%%%%%%%%%%%%%%%%%%%%%%
%%%%%%%%%%%%%%%%%%%%%
%%%%Introduction %%%%%%%%%
%%%%%%%%%%%%%%%%%%%%

\section{Introduction and Summary}\label{first} The
AdS/CFT duality \ct{mal1} relates type IIB string theory on
AdS$_5\times$ S$^5$ with ${\cal N} = 4$ superconformal Yang-Mills
(SYM) theory, and it has been celebrated in the last decade as one
of the exact duality between string and gauge theory. Recently
there has been a lot of works devoted towards the understanding of
the worldvolume dynamics of multiple M2-branes, initiated by
Bagger, Lambert and Gustavsson \cite{Bagger:2006sk} based on the
structure of Lie 3-algebra. In this new development of
understanding of the worldvolume theory of coincident M-branes in
M-theory, a new class of conformal invariant 2+1 dimensional field
theories has been found out. Based on this Aharony, Bergman,
Jafferis and Maldacena (ABJM) \cite{Aharony:2008ug} proposed a new
gauge-string duality between ${\cal N} = 6$ Chern-Simons theory
and type IIA string theory on AdS$_4 \times {\bf CP}^3$. More
precisely,  ABJM theory has been conjectured to be dual to
M-theory on AdS$_4 \times S^7/Z_k$ with $N$ units of four-form
flux which for $k << N << k^5$ can be compactified to type IIA
theory on AdS$_4 \times {\bf CP}^3$, where $k$ is the level of
Chern-Simon theory with gauge group $SU(N)$. This ABJM theory is
weakly coupled for $\lambda << 1$, where $\lambda = N/k$ is the 't
Hooft coupling. Once this duality was proposed, there has been a
numerous effort in understanding the ABJM theory more
\cite{Benna:2008zy}-\cite{Garousi:2008ik}.

In the development of AdS$_5$/CFT$_4$ duality, an interesting
observation is that the ${\cal N}=4$ SYM theory in planar limit
can be described by an integrable spin chain model where the
anomalous dimension of the gauge invariant operators were found
\ct{Beisert:2005tm,zm,intg1,intg2,intg3}. It was further noticed
that the string theory is also integrable in the semiclassical
limit \ct{intg4,tsey3,Okamura:2008jm,Hayashi:2007bq, Chen:2006gea}
and the anomalous dimension of the ${\cal N}=4$ SYM can be derived
from the relation between conserved charges of the rotating string
AdS$_5 \times$ S$^5$. In this connection, Hofman and Maldacena
(HM) \ct{Hofman:2006xt} considered a special limit where the
problem of determining the spectrum in both sides becomes rather
simple. The spectrum consists of an elementary excitation known as
magnon that propagates with a conserved momentum $p$ along the
infinitely long \ct{ik0705,krec,jin,sv1,tsey1,tsey2,tsey3,tsey4,
tsey5,tsey6,tsey7,Spradlin:2006wk,Kalousios:2006xy} or the
finitely long
\ct{Ahn:2008sk,Arutyunov:2006gs,Astolfi:2006is,Astolfi:2007uz}
spin chain. In the dual formulation, the most important ingredient
is the semiclassical string solution, which can be mapped to long
trace operator with large energy and large angular momenta. A more
general class of rotating string solutions are the spiky strings
\cite{Kruczenski:2004wg,krt0607} that describe the higher twist
operators from the field theory view point. Giant magnon solutions
could be thought of as a special limit of such spiky strings with
shorter wavelength. Recently there has been a lot of work devoted
for finding the giant magnon and spike solutions for strings in
more general background, (see for example
\ct{Kluson:2007qu,Bobev:2007bm,Lee:2008sk,ne,Kluson:2008gf}).

The integrability of AdS$_5$/CFT$_4$ in the planar limit using a
Bethe ansatz brings us the hope that the recently proposed
AdS$_4$/CFT$_3$ duality will also be solvable by using a similar
ansatz \ct{Minahan:2008hf}. Indeed, in
\cite{Minahan:2008hf,Gaiotto:2007qi,Gaiotto:2008cg,Gromov:2008qe}
this has been investigated and many interesting results were
found. The magnon solutions were found in the $SU(2)\times SU(2)$
sub-sector of ${\bf CP}^3$. For example the giant magnon found in
\cite{Grignani:2008is,Grignani:2008te} is a solitonic string
living on $R\times S^2 \times S^2$ and rotating uniformly around
the two spheres. In the present paper we would like to find out a
spike solution for the string rotating on $S^2 \times S^2$, and
interpret it as a general class of solution in the worldsheet
theory. We solve the equations of motion and the Virasoro
constraints for the Polyakov action of the string. We write down
the general form of equations of motion which in two different
limits corresponds to the already known giant magnon and the new
spike solution for the string moving in the $SU(2)\times SU(2)$.
The dispersion relations among the various conserved charges have
been found out in both cases. We further study the finite size
corrections to the dispersion relations.

The rest of the paper is organized as follows. In section 2. we
consider a rotating string solution on $R \times S^2 \times S^2$,
which is obtained by fixing some coordinates of AdS$_4\times {\bf
CP}^3$. Taking into account the Polyakov form of the action for
the string in this background, we find the general forms of all
conserved charges. In section 3, we find out the giant magnon and
spike as two different limiting cases and write the dispersion
relation along various conserved charges. For the magnon case, we
reproduce the result obtained in
\ct{Grignani:2008is,Grignani:2008te}. Section 4 is devoted to the
study of finite size effects for both the giant magnon and spike
solutions. In section 5, we present our conclusions.

%%%%%%%%%%%%%%%%%%%%%%%%%%%%%%%
\section{Rotating strings on $R \times S^2 \times S^2$ }\label{third}
%%%%%%%%%%%%%%%%%%%%%%%%%%%%%%%%%

In this section, we will investigate a general class of rotating string
solution on $R\times S^2 \times S^2$ which is a subspace of $AdS_4
\times {\bf CP}^3$. We start by writing down the metric for $AdS_4
\times {\bf CP}^3$ \bea ds^2 &=& \frac{1}{4} R^2 \lb - \cosh^2 \r
\ dt^2 + d\r^2 + \sinh^2 \r \ls d \th^2 + \sin^2 \th d \ph^2 \rs
\rb  \nn && \quad + R^2 \lb d\xi^2 + \cos^2 \xi  \sin^2 \xi \ls d
\ps + \half \cos \th_1 d \ph_1 - \half \cos \th_2 d \ph_2 \rs^2
\rp \nn && \ \ \lp + \frac{1}{4} \cos^2 \xi \ls d \th_1^2 + \sin^2
\th_1 d \ph_1^2 \rs + \frac{1}{4} \sin^2 \xi \ls d \th_2^2 +
\sin^2 \th_2 d \ph_2^2 \rs \rb . \eea While taking $\a^{\pr} =1$,
the curvature radius $R$ is given by $R^2 = 2^{5/2} \pi \l^{1/2}$.
The 't Hooft coupling constant is  $\l \equiv N/k$ where $k$ is
the level of the 3-dimensional ${\cal N}=6 $ ABJM model.

To investigate the string theory dual to spin chain model of SU(2) sector
in the boundary SYM, we first consider the string moving in $R \times S^2 \times S^2$,
which is the subspace of $R \times {\bf CP}^3$ and corresponds to the SU(2)$\times$SU(2)
R-symmetry group of the boundary SYM.
This subspace can be obtained by choosing $\r = 0$, $\ps$ and $\xi =$constant
and then giving the identification $\th_1 = \th_2 \equiv \th$ and $\ph_1 =
\ph_2 \equiv \ph$. Note that this identification reduces $R \times
S^2 \times S^2$ to $R \times S'^2$ effectively, where $S'^2$ can be parameterized by
$\th = \half (\th_1 + \th_2)$ and $\ph = \half (\ph_1 + \ph_2)$ and
corresponds to
the diagonal $SU(2)$ subgroup of $SU(2) \times SU(2)$ R-symmetry.

More precisely, the action for string moving in $R \times {\bf CP}^3$,
where $R$ is the time direction on
AdS$_4$ at $\r = 0$, is
\begin{eqnarray}
S &=& \frac{R^2}{16 \pi} \int d^2 \sigma  \left[
- \partial_{\alpha} t \partial^{\alpha} t + 4 \partial_{\alpha} \xi \partial^{\alpha} \xi
+ 4 \cos^2 \xi \sin^2 \xi \Gamma_{\alpha} \Gamma^{\alpha}  \right. \nonumber \\
&&  \qquad \qquad \qquad + \cos^2 \xi (\partial_{\alpha} \theta_1 \partial^{\alpha} \theta_1
+ \sin^2 \theta_1 \partial_{\alpha} \phi_1 \partial^{\alpha} \phi_1) \nonumber \\
&& \left. \qquad \qquad \qquad + \sin^2 \xi (\partial_{\alpha} \theta_2 \partial^{\alpha} \theta_2
+ \sin^2 \theta_2 \partial_{\alpha} \phi_2 \partial^{\alpha} \phi_2) \right]
\end{eqnarray}
with
\begin{equation}
\Gamma_{\alpha} = \partial_{\alpha} \psi + \frac{1}{2} \cos \theta_1 \partial_{\alpha} \phi_1
- \frac{1}{2} \cos \theta_2 \partial_{\alpha} \phi_2 ,
\end{equation}
where $\alpha, \beta$ implies the string worldsheet indices.
The equations of motion for $\xi$ and $\psi$ are
\begin{eqnarray}    \label{eqcp3}
0 &=& 4 \partial^{\alpha} \partial_{\alpha} \xi - 4 \sin 2 \xi \cos 2 \xi
\Gamma_{\alpha} \Gamma^{\alpha}
+ \sin \xi \cos \xi (\partial_{\alpha} \theta_1 \partial^{\alpha} \theta_1 \nonumber \\
&& + \sin^2 \theta_1 \partial_{\alpha} \phi_1 \partial^{\alpha} \phi_1
- \partial_{\alpha} \theta_2 \partial^{\alpha} \theta_2
- \sin^2 \theta_2 \partial_{\alpha} \phi_2 \partial^{\alpha} \phi_2) ,  \nonumber \\
0 &=& \partial^{\alpha} ( \cos^2 \xi \sin^2 \xi \Gamma_{\alpha} ) \label{pse} .
\end{eqnarray}
When $\psi =$ constant,
$\theta_1 = \theta_2$ and $\phi_1 = \phi_2$ gives $\G_\alpha=0$, which satisfies the
second equation in Eq. (\ref{eqcp3}) and reduces the first equation to a simple form
$0 = 4 \partial^{\alpha} \partial_{\alpha} \xi$.
The simplest solution of this is $\xi =$ constant.
Under these solutions, the open string motion on ${\bf CP}^3$ reduces to
that on $S'^2$ effectively. Therefore,
the rest equations of motion for other fields become
\begin{eqnarray}   \la{metric}
0 &=& \partial^{\alpha} \partial_{\alpha} t , \nonumber \\
0 &=& \partial^{\alpha} \partial_{\alpha} \theta  - 2 \sin \theta \cos \theta
\partial_{\alpha} \phi \partial^{\alpha} \phi, \nonumber \\
0 &=& \partial^{\alpha} \left( \sin^2 \theta \partial_{\alpha} \phi \right) .
\end{eqnarray}
Note that these equations are those for the string sigma model moving on $R \times S^2
\times S^2$ with constraints $\theta_1 = \theta_2 = \theta$ and $\phi_1 = \phi_2 =\phi$.

The reduction from ${\bf CP}^3$ to $S^2 \times S^2$ can be shown with different way using the
complex coordinates. For that, we first consider the embedding $S^7$ to $R^{8}$.
Then, $S^7$ can be described by the constraint equation, in terms of complex coordinates $Z_i$
$( i=1,\cdots, 4)$ or real coordinates $X_a$ $(a=1,\cdots,8)$ on the 8-dimensional Euclidean
space,
\begin{equation}
\frac{R^2}{4}  = \sum_{i=1}^{4} | Z_i |^2 = \sum_{a=1}^{8} X_a^2 ,
\end{equation}
where we set $Z_i = X_i + i X_{i+4}$.
Imposing more constraint
\begin{equation}    \la{concp}
0 = \frac{i}{2} \sum_{i=1}^{4} (Z_i \partial_{\alpha} \bar{Z}_i
- \bar{Z}_i \partial_{\alpha} Z_i ) = \sum_{i=1}^{4} (X_i \partial_{\alpha} X_{i+4}
- X_{i+4} \partial_{\alpha} X_i) ,
\end{equation}
where $\alpha$ implies the world sheet indices, reduces the above $S^7$ to ${\bf CP}^3$
\ct{Grignani:2008te}.
The complex coordinates representing $S^2 \times S^2$ become in terms of the angular
variables in Eq. (\ref{metric}),
\begin{eqnarray}
Z_1 &=& \frac{R}{2} \cos \xi \sin \theta e^{i \phi} , \nonumber \\
Z_2 &=& \frac{R}{2} \cos \xi \cos \theta  , \nonumber \\
Z_3 &=&  \frac{R}{2}  \sin \xi \sin \theta e^{- i \phi} , \nonumber \\
Z_4 &=&  \frac{R}{2}  \sin \xi \cos \theta  ,
\end{eqnarray}
where $\xi$ is a constant. This parameterization
satisfies the constraint for $S^7$ and to satisfy the constraint for ${\bf CP}^3$ in
\eq{concp} we should set $\xi=\frac{\pi}{4}$.  This effectively
describes $S'^2$ as the subspace of ${\bf CP}^3$.

The Polyakov action for a string moving on this $R \times S^2 \times S^2$
is given by
\be
S = \frac{1}{4 \pi} \int d^2 \s \ \sqrt{-\det h} \
h^{\a \b} \pa_{\a} x^{\m} \pa_{\b} x^{\n} G_{\m\n} ,
\ee
with the metric
\bea \la{redmet}
ds^2 &=& \frac{1}{4} R^2 \lb  -dt^2 + \cos^2 \xi \ls d \th_1^2 + \sin^2 \th_1 d \ph_1^2
\rs + \sin^2 \xi \ls d \th_2^2 + \sin^2 \th_2 d \ph_2^2 \rs\rb ,
\eea
which reduces to $R \times S'^2$ under the identification, $\th_1 = \th_2$ and $\ph_1 =
\ph_2$.

%Here, the radius of each sphere $S^2$ depends on $\xi$, so to show the $\xi$
%dependence, we start with the string action on $R \times S^2 \times S^2$ and
%with choosing the conformal gauge $h^{\a\b} = \eta^{\a\b}$.
%Here, we are interested in the string motion moving in $R \times S'^2$, the subspace of
%$R \times S^2 \times S^2$, where $R$ means time direction and $S'^2$ is related to the
%diagonal SU(2)
%R-symmetry group of the boundary gauge theory. For this, we consider the case
%$\rho=0$ only, which implies that the string solution considered here appears
%as a point in AdS space.

In terms of target space coordinates in the conformal gauge $h^{\a\b} = \eta^{\a\b}$,
the effective action on $R \times S'^2$ is written as
\bea
S = \frac{T}{2} \int d^2 \s \lb (\pa_{\ta} t)^2 - (\pa_{\s} t)^2 -
(\pa_{\ta} \th)^2 + (\pa_{\s} \th)^2 - \sin^2 \th \lc (\pa_{\ta}
\ph)^2 - (\pa_{\s} \ph)^2 \rc \rb , \eea where the string tension
$T$ is given by \be \la{tre} T = \frac{\sqrt{2 \l}}{2} \ . \ee To
find the giant magnon or spike solutions for string, we choose the
following parametrization \bea t &=& f(\ta) , \nn \th_1 = \th_2
&=& \th (y), \nn \ph_1 = \ph_2 = \phi &=& \n \ta + g (y) , \eea
where $y = a \ta + b \s$ and $\ta$ and $\s$ run from
$-\infty$ to $\infty$.

Due to the translation symmetry along $t$ and the rotational
symmetry along $\ph_i$'s ($i=1,2$), there exist three conserved
charges and the equations of motion for the corresponding fields
are given by \bea 0 &=& \pa_{\ta}^2 f(\ta) , \nn 0 &=& \pa_y \lb
\sin^2 \th \lc a \n +(a^2 - b^2) g^{\pr} \rc \rb , \eea where
prime implies derivative with respect to $y$. The solutions of
these equations are \bea f(\ta) &=& \k \tau \ , \nn g'(y) &=&
\frac{1}{a^2 -b^2} \frac{c - a\n \sin^2 \th }{\sin^2 \th} \ , \eea
where $\k$ and $c$ are integration constants. The equation of
motion for the world sheet metric $h^{\a\b}$, gives rise to the
Virasoro constraints $T_{\a\b} = 0$ where \bea T_{\a\b} &\equiv&
\frac{1}{\sqrt{- \det h}} \frac{\pa {\cal L}}{\pa h^{\a\b}} \nn
&=& \frac{T}{2} \ls \pa_{\a} x^{\m} \pa_{\b} x^{\n} G_{\m\n} -
\half \eta_{\a\b} \eta^{\a' \b'}\pa_{\a'} x^{\m} \pa_{\b'} x^{\n}
G_{\m\n}  \rs . \eea Due to the symmetric property of the metric,
the independent constraints are three, $T_{\ta\ta}$, $T_{\ta\s}$
and $T_{\s\s}$. Furthermore, the conformal nature of the Polyakov
action gives rise to the relation $T_{\ta\ta} = T_{\s\s}$, so only two of them are
independent. For later
convenience, these two Virasoro constraints are rewritten as \bea
0 &=&  T_{\ta\ta} + T_{\s \s} + 2 T_{\ta\s} \ , \nn 0 &=&
T_{\ta\ta} + T_{\s \s} - \frac{a^2 + b^2}{a b} T_{\ta\s} .
\label{vira} \eea The first line of \eq{vira} gives a
first order differential equation for $\th$, \be \la{the} \th' =
\frac{b \n}{a^2 -b^2} \frac{\sqrt{ (\sin^2 \th_{max} - \sin^2 \th)
(\sin^2 \th - \sin^2 \th_{min} )}}{\sin \th} , \ee where
$\sin \th_{max}$ and $\sin \th_{min}$ satisfy
\bea
\la{inf} \sin^2 \th_{max} + \sin^2 \th_{min} &=& \frac{\k^2
(a-b)^2 + 2b\n c}{b^2 \n^2} ,\nn \sin^2 \th_{max} \cdot \sin^2
\th_{min} &=& \frac{c^2}{b^2 \n^2} . \eea The second line of \eq{vira}
reduced to a relation among various constants. From
this, one can find \be \la{con1} a = \frac{\n}{\k^2} c . \ee
Using this, we finally obtain
\be \la{the2}
\th' =
\frac{b \n}{a^2 -b^2} \frac{\sqrt{ (\frac{c^2}{\k^2 b^2} - \sin^2 \th)
(\sin^2 \th - \frac{\k^2}{\n^2} )}}{\sin \th} .
\ee

Now let us proceed to write various conserved charges for this
system. The energy is given by \bea E &\equiv& T \int d \s
\pa_{\ta} t \nn &=& 2 T \int_{\th_{min}}^{\th_{max}} d \th  \
\frac{\k (a^2 -b^2)}{b^2 \n} \frac{\sin \th}{\sqrt{ (\sin^2
\th_{max} - \sin^2 \th) (\sin^2 \th - \sin^2 \th_{min} )}} \eea
and two angular momenta are
\bea
J_1 &\equiv& - T  \cos^2 \xi \int d \s\sin^2 \th \pa_{\ta} \ph \nn
&=& -2 T \int_{\th_{min}}^{\th_{max}} d \th \frac{1}{2 b^2 \n}
      \frac{\sin \th (ac - b^2 \n \sin^2 \th)}{\sqrt{ (\sin^2 \th_{max} - \sin^2 \th)
(\sin^2 \th - \sin^2 \th_{min} )}} , \nn
J_2 &\equiv& - T \sin^2 \xi \int d \s  \sin^2 \th \pa_{\ta} \ph \nn
      &=& -2 T \int_{\th_{min}}^{\th_{max}} d \th \frac{1}{2 b^2 \n}
      \frac{\sin \th (ac - b^2 \n \sin^2 \th)}{\sqrt{ (\sin^2 \th_{max} - \sin^2 \th)
(\sin^2 \th - \sin^2 \th_{min} )}} .
\eea
Note that $J_1$ and $J_2$ are angular momentum on each sphere $S^2$.
To consider a giant magnon or spike
solution, we have to define the world sheet momentum $p$, which is
identified with the angle difference $\D \ph \equiv p$,
%\footnote{The azimuthal angles of the two S$^2$'s
%are chosen in such a way that $\Delta \phi_1 \equiv p$ and $\Delta \phi_2 \equiv - p$}
\bea \la{ang} \D \ph &\equiv& - \int d \ph = - 2
\int_{\th_{min}}^{\th_{max}} d \th \frac{g'}{\th'} \nn &=& - 2
\int_{\th_{min}}^{\th_{max}} d \th \frac{1 }{b^2 \n}
      \frac{(b c - a b \n \sin^2 \th)}{\sin \th \sqrt{ (\sin^2 \th_{max} - \sin^2 \th)
(\sin^2 \th - \sin^2 \th_{min} )}}  ,
\eea
where we use a minus sign for making the angle difference positive.

\section{Giant magnon and Spike solutions}

By using various quantities defined in the previous section, we
now proceed to find the relation among various conserved charges.
Before doing this, first we choose the infinite size limit, which
implies infinite angular momentum in case of giant magnon and
infinite angle between two end points of a spike. Note that this
infinite size limit can be described by setting $\sin \th_{max} =
1$ in both cases. In this case, \eq{the} is reduced
to \be \th' = \frac{b \n}{a^2 -b^2} \frac{\cos \th \sqrt{ (\sin^2
\th - \sin^2 \th_{min} )}}{\sin \th} . \ee Due to the cosine term
in the above equation, all the conserved charges diverge except
for a special region of parameters. Below, we will investigate the
solutions in this region.

%In this limit, the second equation in \eq{inf} gives the value of
%$\sin^2 \th_{min}$ as \be \sin^2 \th_{min} = \frac{c^2}{b^2 \n^2}
%. \ee

%The first equation in \eq{inf} gives rise to another
%relation among the parameters. Using this along with \eq{con1},
%one can fix $b$ as \be \la{con2} b = \k c . \ee

\subsection{Giant magnon solution}

In this section, we will consider a magnon solution which has infinite charges,
$E$ and $J \equiv J_1 + J_2$, but finite difference of them, $E-J$.
For this, we should choose from \eq{the2}
\be
\sin \th_{max} = - \frac{\k}{\n} ,
\ee
where the minus sign is inserted for later consistency.
In the infinite size limit $\n$ can be rewritten as
\be \la{con2}
\n =-\k ,
\ee
where we assume that $\k$ is positive.
Using \eq{con1} and
\eq{con2}, $E-J$ is given by \be \la{rEJ} E - J = 2 T
\int_{\th_{min}}^{\pi/2} d \th
      \frac{ \sin \th  \cos \th }
      {\sqrt{ (\sin^2 \th - \sin^2 \th_{min} )}} = 2 T \sqrt{1 - \sin^2 \th_{min} } ,
\ee
where $\sin \th_{min} = c/\k b$.
Note that $E-J$ has no divergence like our expectation.
The value of the world sheet momentum  $p$ corresponding to the angle difference is
also obtained using \eq{con1} and \eq{con2}
\be
\D \ph = 2  \int_{\th_{min}}^{\pi/2} d \th  \frac{c }{\k b }
      \frac{\cos \th}{\sin \th \sqrt{(\sin^2 \th - \sin^2 \th_{min} )}}
       = 2 \arccos (\sin \th_{min}) .
\ee
Finally, we
obtain the dispersion relation for a giant magnon as
\be \la{dma}
E-J = \sqrt{2 \l} \left| \sin \frac{p}{2} \right| ,
\ee where we
replace the string tension $T$ with the 't Hooft coupling $\l$.
This is the dispersion relation for an open string rotating in
$S'^2$  effectively, which is dual to the open spin chain
in the SU(2) diagonal R-symmetry subgroup.

In Ref. \cite{Gaiotto:2008cg,Grignani:2008is}, a single trace operator
corresponding to a closed spin chain in $SU(2) \times
SU(2)$ sector is considered. Moreover, it was shown that the dual
string solution for this closed spin chain is a closed string rotating in
$S^2 \times S^2$, which is a combination of two open strings, each
rotating on different $S^2$.
In Ref. \cite{Gaiotto:2008cg,Grignani:2008is,Grignani:2008te},
the open string corresponding to half of the closed string has also the
same dispersion relation
in \eq{dma} but the angular momentum $J$ is given by $J=J_1$ or $J_2$.
However, in this case
the giant magnon describes not the open spin chain in
the diagonal $SU(2)$  but that in one of the $SU(2)$
inside $SU(2) \times SU(2)$ R-symmetry group.

\subsection{Spike solution}

To find a spike solution, we should impose that $J$ is finite.
For this, we choose $\sin \th_{max} = \frac{c}{\k b}$ in \eq{the2}.
In the infinite size limit, $\k$ can be rewritten as in terms of $c$ and $b$
\be
\k = \frac{c}{b} .
\ee
Then, $\sin \th_{min}$ becomes
\be
\sin \th_{min} = \frac{c}{b \n}
\ee
Using these, $E-T \D \ph$ and $J$ become \bea    \la{presp} E - T
\D \ph &=& 2 T \int_{\th_{min}}^{\pi/2} d \th
      \frac{ \sin \th_{min} \cos \th}
      { \sin \th \sqrt{ (\sin^2 \th - \sin^2 \th_{min} )}}
      = 2 T \arccos (\sin \th_{min}) , \nn
J &=& - 2 T \int_{\th_{min}}^{\pi/2} d \th
      \frac{ \sin \th \cos \th}
      {\sqrt{ (\sin^2 \th - \sin^2 \th_{min} )}}
      = - 2 T \sqrt{1 - \sin^2 \th_{min} } .
\eea  Notice that if the orientation
of the rotation in $\ph_i$-direction $(i=1,2)$, is changed, then
we can obtain a positive $J$. From now on, we consider a positive
$J$. Then, $E-T \D \ph$ and $J$ can be rewritten in terms of new
variable $\td{\th} = \pi/2 - \th_{min}$ as \bea E - T \D \ph &=&
\sqrt{2 \l} \td{\th} , \nn J &=& \sqrt{2 \l} \sin \td{\th} , \eea
with $J=J_1+J_2$ \bea J_1 &=&  \frac{\sqrt{2 \l}}{2} \sin
\td{\th} ,\nn J_2 &=&  \frac{\sqrt{2 \l}}{2} \sin \td{\th} .
\eea
%For $\xi=0$ or $=\pi/2$, this spike solution reduces to one
%in $R \times S^2$.

\section{Finite Size effects}

In the previous section, we found a giant magnon and a spike
solution in the infinite size limit. Here, we will investigate the
finite size effect on them \footnote{Finite size effect for the
membrane in AdS$_4\times$ S$^7$ has been investigated in Ref.
\cite{Ahn:2008gd}.}. To do so, we have to investigate the
solitonic string solution when $\th_{max} \ne \pi/2$.

\vskip .1in \noindent {\bf Giant magnon case:} \\
For the magnon case, $\sin \th_{min}$ and $\sin \th_{max}$ become
\bea
\sin \th_{max} &=& - \frac{\k}{\n} , \nn \sin \th_{min} &=& \
\frac{c}{\k b} ,
\eea
For the simple calculation, we replace the
variable $\th$ to $z \equiv \cos \th$. With this new variable $z$,
the conserved charges are rewritten  as \bea    \la{deeq} E &=& 2
T \ \frac{z_{max}^2 - z_{min}^2}{z_{max} \sqrt{1-z_{min}^2}} \
K(x), \nn J &=& 2 T z_{max} \  \lb  K(x) - E(x) \rb   , \nn
\frac{\D \ph}{2} &=& \frac{\sqrt{1-z_{min}^2}}{z_{max}
\sqrt{1-z_{max}^2}} \Pi \ls \frac{z_{max}^2 -
z_{min}^2}{\sqrt{z_{max}^2-1}};x \rs
 - \frac{\sqrt{1-z_{max}^2}}{z_{max} \sqrt{1-z_{min}^2}}  K(x),
\eea by using the elliptic integrals of the first, second and
third kinds \bea K(x) &=&  \int_{z_{min}}^{z_{max}} dz
\frac{z_{max}}{\sqrt{(z_{max}^2 - z^2)(z^2 - z_{min}^2)}}  , \nn
E(x) &=&  \int_{z_{min}}^{z_{max}} dz \frac{z^2}{z_{max}
\sqrt{(z_{max}^2 - z^2)(z^2 - z_{min}^2)}} , \nn \Pi \ls
\frac{z_{max}^2 - z_{min}^2}{\sqrt{z_{max}^2-1}};x \rs &=&
 \int_{z_{min}}^{z_{max}} dz
\frac{ z_{max} (1-z_{max}^2)}{(1-z^2) \sqrt{(z_{max}^2 - z^2)(z^2
- z_{min}^2) }} , \eea where $z_{max}^2 \equiv \cos^2 \th_{min}=
\frac{k^2 b^2 -c^2}{\k^2 b^2}$, $z_{min}^2 \equiv \cos^2 \th_{max}
= \frac{\n^2 - \k^2}{\n^2}$ and $x = \sqrt{1-
\frac{z_{min}^2}{z_{max}^2}}$. The expansion of
the conserved charges to ${\cal O} (z_{min}^2)$ and
${\cal O} (z_{max}^2)$, gives rise to \bea    \la{fe1} E-J &\approx&
2T \ls \left| \sin \frac{p}{2} \right| - \frac{z_{max}
z_{min}^2}{4}\rs . \eea

The leading behaviors of $E$ and $z_{max}$ are given by \bea
\la{ere} E &\approx&  2 T z_{max} \log \frac{4 z_{max}}{z_{min}} ,
\nn z_{max} &\approx&  \left| \ \sin \frac{p}{2} \right| . \eea
Using these relations, we finally obtain the approximate form of
the dispersion relation for a magnon with the finite size
correction \bea    \la{fco} E - J &=& 2 T \left(\left|  \sin
\frac{p}{2} \right| - 4 \left| \sin^3 \frac{p}{2} \right| e^{- E /
( T \left| \sin \frac{p}{2} \right| )}\right) \nn &=& \sqrt{2 \l}
\left(\left|  \sin \frac{p}{2} \right| - 4 \left| \sin^3
\frac{p}{2} \right| e^{- 2 E / ( \sqrt{2 \l} \left|  \sin
\frac{p}{2} \right| )}\right) . \eea For the infinite size case $E
\ {\rm and} \ J \to \infty$, this gives the same result obtained in the
previous section. The second term in the above equation is the
finite size correction. Note that this result is the finite
correction for a giant magnon dual to
open spin chain and that this correction corresponds to half of that for the
closed string \cite{Grignani:2008te}.

\vskip.1in \noindent
{\bf Spike case:}\\
In this section, we will calculate the finite size effect for
a spike. Note that as previously mentioned, for
considering a positive angular momentum, we should consider the
angular momentum for a spike as $J' \equiv - J$ by changing the
directions of rotation. Keeping this in mind, we now start to
calculate the finite size effect for a spike.
For the spike, $\sin \th_{min}$ and $\sin \th_{max}$ are given by
\bea    \la{parsp}
\sin \th_{min} &\equiv&  \sqrt{1-z_{max}^2} =  \frac{\k}{\n}  , \nn
\sin \th_{max} &\equiv&  \sqrt{1-z_{min}^2} =  \frac{c}{\k b}  .
\eea
Note that in the infinite size limit, this parameterization reduces to
the one used in the previous section, $\k = \frac{c}{b}$ and
$\sin \th_{min} = \frac{c}{b\n} $.

Using these, the conserved charges can be rewritten as
\bea
E &=& 2 T  \ \frac{  z_{max}^2 - z_{min}^2  }{z_{max} \sqrt{1-z_{max}^2}} \ K(x)  ,\nn
J' &=& 2 T \ \frac{1}{z_{max}} \ls  z_{max}^2 E (x) - z_{min}^2 K(x) \rs , \nn
\frac{\D \ph}{2} &=& \frac{\sqrt{1-z_{min}^2}}{z_{max} \sqrt{1-z_{max}^2}} \lb K(x)
- \Pi \ls \frac{z_{max}^2 - z_{min}^2}{\sqrt{z_{max}^2-1}};x \rs \rb .
\eea

Here, the angular momentum $J'$ is given by \be   \la{jeq} J'
\approx 2 T z_{max}  - \ls \half + \log \frac{4 z_{max}}{z_{min}}
\rs \frac{T z_{min}^2}{z_{max}}  , \ee at ${\cal O} (z_{min}^2)$.
Note that for the infinite size limit, $z_{min} \to 0$, the second
term in the right hand side vanishes, so $J'$ is always finite as
it should be. The dispersion relation for a spike $E- \D \ph$ is
given by up to ${\cal O}(z_{min}^3)$ and ${\cal O}(z_{max}^3)$ as
\bea    \la{predis} E- T \D \ph &\approx& 2 T \arcsin z_{max} \nn
  && - \lb  \ls \frac{1}{2 z_{max}} - \frac{ z_{max}}{4} \rs T
  + \ls \frac{1}{z_{max}} + \frac{z_{max}}{2} \rs T \log \frac{4 z_{max}}{z_{min}} \rb z_{min}^2 .
\eea
Using \eq{jeq}, $2 T \arcsin z_{max}$ can be approximately rewritten as
\be
2 T \arcsin z_{max} \approx 2 T \arcsin \frac{J'}{2T}
+ \frac{\half + \log \frac{4 z_{max}}{z_{min}}}{ z_{max} \sqrt{1-\frac{J'^2}{4 T^2} }}
\   T z_{min}^2 .
\ee
To rewrite the dispersion relation in terms of the physical quantities,
$z_{min}$ and $z_{max}$ should be replaced with $E$ and $J'$. From the leading term
of $E$
we obtain
\be
z_{min} = 4 z_{max} e^{-E/2T z_{max}}
\ee
and the leading term of $J'$ gives
\be
z_{max} = \frac{J'}{2 T} .
\ee
Using these, we finally obtain the dispersion relation for a finite size spike solution
\bea    \la{edis}
E- T \D \ph &\approx& 2 T \arcsin \frac{J'}{2T}
- \lb \ls 4 - \frac{4}{\sqrt{1- (J'/2T)^2}} - \frac{J'^2}{2T^2} \rs J' \rp \nn
&& \lp  + \ls 8 - \frac{8}{\sqrt{1- (J'/2T)^2}}
+ \frac{J'^2}{T^2} \rs E \rb \ e^{-2E/J'} .
\eea
In the infinite size limit, since the spike has infinite $E$ and $\D \ph$
with the finite $J'$, the above result gives rise to
the same dispersion relation obtained in previous section
\be
E- T \D \ph \approx 2 T \arcsin z_{max} =  2 T \ls \frac{\pi}{2} - \th_{min} \rs .
\ee
The second term in \eq{edis} corresponds to the finite size
effect for a spike.
For $z_{max} << 1$ ($T >> J'$), the above dispersion relation reduces to
\be
E- T \D \ph \approx 2 T \arcsin \frac{J'}{2T} +  \ls 1 + \frac{3J' E}{8T^2}  \rs
\ \frac{J'^3}{T^2}\ e^{-2E/J'} ,
\ee
where $T = \frac{\sqrt{2 \l}}{2}$.

\section{Discussions} We have studied, in this paper, the rotating
string in the diagonal $SU(2)$ inside AdS$_4 \times {\bf CP}^3$.
We have solved the most general form of the equations of motion of
the rotating string on $R\times S^2 \times S^2$, and have found
out the most general form of all conserved charges. We have shown
the existence of both the already known giant magnon, and the new
spike solutions for the string and have found out the relevant
dispersion relation among various charges in the infinite size
limit. Furthermore, we have studied the finite size correction in
both cases. It will be interesting to find out a three spin giant
magnon with one spin along the AdS$_4$ and two angular momenta one
in each of S$^2$ and study the dual gauge theory. Another
interesting aspect will be to write down the semiclassical
scattering of the giant magnon and spike solutions on AdS$_4
\times {\bf CP}^3$. We wish to come back to some of these issues
in future.

\vskip .2in \noindent {\bf Acknowledgements:} This work was
supported by the Science Research Center Program of the
Korean Science and Engineering Foundation through the Center for
Quantum SpaceTime (CQUeST) of Sogang University with grant number
R11-2005-021.
C. Park was partially supported by the Korea Research Council of Fundamental Science and Technology (KRCF).

%%%%%%%%%%%%%%%%%%%%%%%%%%%%%%%%%%%%%%
%%%%%%% Thebibligraphy %%%%%%%%%%%%%%%%%%%%%
%%%%%%%%%%%%%%%%%%%%%%%%%%%%%%%%%%%%%

%%%%%%%%%%%%%%%%% Journal Macros %%%%%%%%%%%%%%%%%%%%%%%%%%%
\newcommand{\np}[3]{Nucl. Phys. {\bf B#1}, #2 (#3)}
\newcommand{\pprd}[3]{Phys. Rev. {\bf D#1}, #2 (#3)}
\newcommand{\jjhep}[3]{J. High Energy Phys. {\bf #1}, #2 (#3)}
%\newcommand{\hep}[1]{{\tt hep-th/{#1}}}
%\newcommand{\app}[3]{Ann. Phys. {\bf #1}, #2, (#3)}
%\newcommand{\prp}[3]{Phys. Rept. {\bf #1}, #2, (#3)}
%\newcommand{\jmp}[3]{J. Math. Phys. {\bf #1}, #2, (#3)}
%%%%%%%%%%%%%%%%%%%%%%%%%%%%%%%%%%%%%%%%%%%%%%%%%%%%%%%%%%%%%%

\end{document}